\documentclass[preprint,3p,11pt,authoryear]{elsarticle}

\usepackage{booktabs}
\usepackage{amsmath}
\usepackage{amssymb}
\usepackage{lineno}
\usepackage{color}
\usepackage{url}
\usepackage{bm}

\journal{arXiv}

\begin{document}

\begin{frontmatter}

\noindent \textcolor{blue}{
\copyright ~ 2024 American Society of Civil Engineers.
This manuscript has been accepted for publication in \textit{ASCE-ASME Journal of Risk and Uncertainty in Engineering Systems, Part A: Civil Engineering} and is already published.
The DOI is \url{https://doi.org/10.1061/AJRUA6.RUENG-1314}.
Please cite this article as follows. T. Yaoyama, T. Itoi, J. Iyama (2024) Stress Resultant--Based Approach to Mass Assumption--Free Bayesian Model Updating of Frame Structures, \textit{ASCE-ASME Journal of Risk and Uncertainty in Engineering Systems, Part A: Civil Engineering}, 10(4): 04024055.}

\title{Stress Resultant--Based Approach to Mass Assumption--Free Bayesian Model Updating of Frame Structures}

\author[label1]{Taro Yaoyama\corref{cor1}}\ead{yaoyama@g.ecc.u-tokyo.ac.jp}
\author[label1]{Tatsuya Itoi}
\author[label1]{Jun Iyama}

\cortext[cor1]{corresponding author}

\affiliation[label1]{
    organization={Department of Architecture, Graduate School of Engineering, The University of Tokyo},
    addressline={7-3-1, Hongo}, 
    city={Bunkyo-Ku},
    postcode={113-8656}, 
    state={Tokyo},
    country={Japan}
}

\begin{abstract}
Bayesian model updating facilitates the calibration of analytical models based on observations and the quantification of uncertainties in model parameters such as stiffness and mass. This process significantly enhances damage assessment and response predictions in existing civil structures.
Predominantly, current methods employ modal properties identified from acceleration measurements to evaluate the likelihood of the model parameters. 
This modal analysis-based likelihood generally involves a prior assumption regarding the mass parameters.
\textcolor{black}{
In civil structures, accurate determination of mass parameters proves challenging owing to the significant uncertainty and time-varying nature of imposed loads.
}
The resulting inaccuracy potentially introduces biases while estimating the stiffness parameters, which affects the assessment of structural response and associated damage.
Addressing this issue, the present study introduces a stress-resultant-based approach for Bayesian model updating independent of mass assumptions.
This approach utilizes system identification on strain and acceleration measurements to establish the relationship between nodal displacements and elemental stress resultants.
Employing static analysis to depict this relationship aids in assessing the likelihood of stiffness parameters.
Integrating this static-analysis-based likelihood with a modal-analysis-based likelihood facilitates the simultaneous estimation of mass and stiffness parameters.
The proposed approach was validated using numerical examples on a planar frame and experimental studies on a full-scale moment-resisting steel frame structure.
\end{abstract}

\end{frontmatter}

\section{Introduction}

Vibration-based structural health monitoring (SHM) has emerged as a prominent tool in civil engineering for evaluating the actual performance of existing civil structures.
It can be classified into four levels \citep{rytterA1993,simoenE2015,wordenK2015}: (1) \textit{detection}, determining the presence of damage; (2) \textit{localization}, identifying the damage location; (3) \textit{quantification}, ascertaining the damage extent; and (4) \textit{prediction}, evaluating the response and performance for future natural hazards.
Finite element (FE) model updating shows substantial potential across these levels, from detection to prediction, by calibrating analytical models to correlate simulated responses with observed structural behavior.
However, FE model updating often encounters challenges, including ill-posed and ill-conditioned inverse problems caused by measurement noise, modeling errors, and data scarcity. These challenges introduce significant uncertainty in estimated parameters \citep{beckJL1998,simoenE2015}.
Bayesian FE model updating offers a solution \citep{beckJL1998,vanikMW2000,beckJL2001,beckJL2002,chingJ2006,beckJL2010,behmaneshI2015}, providing a rigorous probabilistic framework that integrates prior knowledge with observed data to determine structural parameters with quantified uncertainty.

Diverse methodologies for FE model updating and SHM utilize modal properties (e.g., natural frequencies and mode shapes) extracted from acceleration measurements \citep{chingJ2006,yinT2017,behmaneshI2018,lamHF2018,huangY2019a,dasA2021,yinT2022}. The objective function (or the likelihood function in the Bayesian context) is based on the divergence between observed modal properties and those simulated through modal analysis, using specified mass and stiffness parameters. However, acceleration-based methods exhibit limitations.
Primarily, acceleration responses are indicative of the global, rather than local, attributes of a structure, thus limiting their utility in detecting localized structural damages, such as buckling or cracks \citep{yaoGC1992,beckJL2001,singhMP2018}. 
Furthermore, these methods mainly focus on stiffness, requiring prior knowledge or assumptions regarding mass. This requisite arises from the challenge in simultaneously identifying mass and stiffness, considering the infinite permutations of these parameters that can satisfy an eigenvalue problem for designated modal properties \citep{beckJL2002,zengJ2022}.

\textcolor{black}{
Accurate mass determination in civil structures presents a formidable challenge owing to uncertain or time-varying imposed loads such as the weights of occupants, vehicles, contained assets, and equipment \citep{ikedaY2008,xuB2019}. 
}
This inaccuracy may introduce an undesirable bias in stiffness estimation and subsequent damage assessment. Various studies have attempted to mitigate this issue. \citet{ikedaY2008} developed a deterministic method for mass identification from mode shapes using linear programming.
\citet{xuB2019} introduced a model-free technique for identifying mass, stiffness, and nonlinear restoring force by integrating the extended Kalman filter (EKF), unscented Kalman filter (UKF), and Chebyshev polynomial. \citet{zengJ2022} proposed a Bayesian approach for the simultaneous identification of mass and stiffness, although it required known added mass. However, the aforementioned studies were confined to applications in shear-type multi-degree-of-freedom (MDOF) structures and did not address localized damage identification.

Conversely, dynamic strain measurements exhibit notable sensitivity to elemental stiffness, thereby facilitating enhanced damage localization \citep{esfandiariA2010,esfandiariA2014,pedramM2016,singhMP2018,matarazzoTJ2018}. 
Several SHM studies explored the combined utilization of strain and acceleration measurements. Although primarily focused on damage assessments rather than model updating, \citet{iyamaJ2021} and \citet{iyamaJ2023} transformed acceleration and strain measurements into mode shapes in terms of displacement and resultant stress (axial force and bending moment), thereby establishing a relationship between external deformation and internal force.
This relationship explicitly yields elemental stiffness without any reliance on mass (external load) assumptions.
\citet{yaoyamaT2023} subsequently employed this relationship in model updating and damage localization of steel frame structures; however, they entailed a deterministic approach and did not address mass estimation.

The present study proposes a stress-resultant-based methodology for Bayesian model updating of existing civil structures without any mass assumptions. This approach aims to enhance post-event structural damage detection, localization, and assessment as well as pre-event response predictions. It applies system identification to both strain and acceleration measurements to discern the relationship between nodal displacements and elemental stress resultants in the modal space.
This relationship is then leveraged to establish the likelihood of stiffness parameters using static analysis, thereby eliminating the need for explicit mass parameters. Integrating such a static-analysis-based likelihood with traditional modal-analysis-based likelihood facilitates the simultaneous estimation of mass and stiffness parameters. The proposed methodology was validated using numerical examples on a planar frame and experimental studies on a full-scale moment-resisting steel frame structure.

\section{STRESS-RESULTANT-BASED APPROACH TO BAYESIAN MODEL UPDATING}

\subsection{Structural Model Description}

\textcolor{black}{
Consider an $N$-degree-of-freedom ($N$-DOF) finite element model composed of $M$ structural elements.
Let $\bm{p} \in \mathbb{R}^N$ be a nodal load vector and $\bm{d} \in \mathbb{R}^N$ be a nodal displacement vector,
and $\mathbf{K}(\bm{\theta}_\mathrm{K}) \in \mathbb{R}^{N \times N}$ be a (global) stiffness matrix parameterized by uncertain stiffness parameters $\bm{\theta}_\mathrm{K}$, i.e., $\bm{p} = \mathbf{K}(\bm{\theta}_\mathrm{K}) \bm{d}$.
For individual elements $m = 1,...,M$, let ${}_m\bm{r} \in \mathbb{R}^{N_m}$ be an elemental stress-resultant vector that contains all stress-resultant components of the $m$th structural element, and ${}_m\bm{e} \in \mathbb{R}^{N_m}$ be the corresponding deformation vector.
For a two-dimensional (2D) beam element, ${}_m \bm{r} \in \mathbb{R}^3$ has one component of axial force and two components of bending moment at both beam-ends and ${}_m \bm{e} \in \mathbb{R}^3$ has one component of axial deformation and two components of rotational deformation at both beam-ends, although it may vary by definition.
The two vectors ${}_m\bm{r}, {}_m\bm{e}$ are related through an elemental stiffness matrix ${}_m\mathbf{K} \in \mathbb{R}^{N_m \times N_m}$, that is, ${}_m\bm{r} = {}_m\mathbf{K}~{}_m\bm{e}$.
Define the vectors that concatenate all elemental stress-resultant and deformation components in a structure as $\bm{r} = \{ {}_1\bm{r}^\top, ..., {}_{N_\mathrm{ele}}\bm{r}^\top \}^\top \in \mathbb{R}^M$ and $\bm{e} = \{ {}_1\bm{e}^\top, ..., {}_{N_\mathrm{ele}}\bm{e}^\top \}^\top \in \mathbb{R}^M$, where $N_\mathrm{ele}$ is the number of elements in the structure and $M = \sum_{m = 1}^{N_\mathrm{ele}} N_m$.
}

\textcolor{black}{
The equilibrium condition of the structure gives \citep{livesleyRK1975}
\begin{linenomath*}\begin{align}
    \bm{p} = \mathbf{H} \bm{r}
\end{align}\end{linenomath*}
where $\mathbf{H} \in \mathbb{R}^{N \times M}$ represents an equilibrium matrix, which describes the connectivity between nodes and elements.
Assuming elastic and infinitesimal deformations, the principle of virtual work yields the following compatibility condition,
\begin{linenomath*}\begin{align}
    \bm{e} = \mathbf{H}^\top \bm{d}
\end{align}\end{linenomath*}
Therefore, elemental stress-resultant can be associated with nodal displacement as follows
\begin{linenomath*}\begin{align}
    \bm{r}
    &= \mathbf{K}_\mathrm{m}(\bm{\theta}_\mathrm{K}) ~ \bm{e} \nonumber \\
    &= \mathbf{K}_\mathrm{m}(\bm{\theta}_\mathrm{K}) ~ \mathbf{H}^\top \bm{d} \label{eq:stressres}
\end{align}\end{linenomath*}
where $\mathbf{K}_\mathrm{m}(\bm{\theta}_\mathrm{K}) \in \mathbb{R}^{M \times M}$ represents a stiffness matrix for an ``unassembled'' structure, i.e.,}
\begin{linenomath*}\begin{align}
\textcolor{black}{
    \mathbf{K}_\mathrm{m}(\bm{\theta}_\mathrm{K}) =
    \begin{bmatrix}
        {}_1\mathbf{K} & ~              & ~          & \mathbf{0}                    \\
        ~              & {}_2\mathbf{K} & ~          & ~                             \\
        ~              & ~              & \ddots     & ~                             \\
        \mathbf{0}     & ~              & ~          & {}_{N_\mathrm{ele}}\mathbf{K} \\
    \end{bmatrix}
}
\end{align}\end{linenomath*}

\textcolor{black}{For dynamic loading, consider the following eigenvalue problem:
\begin{linenomath*}\begin{align} \label{eq:eigen}
    \left(\mathbf{K}(\bm{\theta}_\mathrm{K}) - \omega_k^2 \mathbf{M}(\bm{\theta}_\mathrm{M})\right) \bm{d}_k = \mathbf{0}
\end{align}\end{linenomath*}
where $\mathbf{M}(\bm{\theta}_\mathrm{M}) \in \mathbb{R}^{N\times N}$ represents the (global) mass matrix parameterized by uncertain mass parameters $\bm{\theta}_\mathrm{M}$.
$\omega_k \in \mathbb{R}$ and $\bm{d}_k \in \mathbb{R}^N$ represent the $k$th natural angular frequency and displacement mode shape, respectively.
Using Eq.~(\ref{eq:stressres}), the corresponding stress-resultant mode shape is introduced as $\bm{r}_k = \mathbf{K}_\mathrm{m}(\bm{\theta}_\mathrm{K}) \mathbf{H}^\top \bm{d}_k$.
For consistency in expressions, $\bm{d}_k$ and $\bm{r}_k$ are termed \textit{modal displacement (MD)} and \textit{modal stress-resultant (MSR)} (similarly, \textit{modal acceleration (MA)}, \textit{modal strain (MS)}, and \textit{modal bending moment (MBM)} are defined later).
}

\subsection{Bayesian Model Updating}

\textcolor{black}{
Let $\mathcal{D} = \{ \bar{\omega}_k, \bar{\bm{d}_k}, \bar{\bm{r}}_k \}_{k=1}^K$ denote a dataset of observed modal properties,
where $\bar{\omega}_k \in \mathbb{R}$, $\bar{\bm{d}_k} \in \mathbb{R}^{N_\mathrm{obs}}$, and $\bar{\bm{r}}_k\in \mathbb{R}^{M_\mathrm{obs}}$ represent observations for the $k$th natural frequency, MD vector, and MSR vector, respectively.
$N_\mathrm{obs}$ and $M_\mathrm{obs}$ represent the numbers of observed components for nodal displacement and elemental stress resultant.
$\bar{\bm{d}}_k$ is normalized such that $\|\bar{\bm{d}}_k\|_2 = 1$, where $\|.\|_2$ represents an operator for the $L_2$ norm of a vector.
Accordingly, $\bar{\bm{r}}_k$ is normalized such that the relationship between $\bar{\bm{d}}_k$ and $\bar{\bm{r}}_k$ is preserved;
i.e., the raw data of both MD and MSR are divided by the $L_2$ norm of the raw MD. 
The concatenated vectors, $\bar{\bm{\omega}} = \{\bar{\omega}_1, ..., \bar{\omega}_K \}^\top$,
$\bar{\bm{d}} = \{ \bar{\bm{d}}_1^\top, ..., \bar{\bm{d}}_K^\top \}^\top$, and $\bar{\bm{r}} = \{ \bar{\bm{r}}_1^\top, ..., \bar{\bm{r}}_K^\top \}^\top$, are also defined.
These observed modal properties are assumed to be deterministically obtained by applying appropriate modal identification techniques simultaneously to acceleration and dynamic strain measurements.
Specifically, mode shapes in terms of dynamic strain are converted into MSR (i.e., mode shapes in terms of axial force and bending moment) by exploiting a prior knowledge of the material and geometric properties of the cross-sections at which strain gauges are installed.
Details on such conversion are described in the EXPERIMENTAL VALIDATION section.
Herein, we assume that all primary nodal DOFs subjected to certain external forces are measured in terms of acceleration, whereas the number of observed stress-resultant components is not constrained.
For notation simplicity, the following formulation assumes a scenario in which only a single observation is obtained; therefore, $K$ is equal to the number of identified vibration modes.
However, this formulation can be simply extended to scenarios with multiple observations.
}

\textcolor{black}{
As in most existing modal-analysis-based Bayesian methods (e.g., \citet{vanikMW2000}, \citet{beckJL2001}), the following generative process of the modal data $\{\bar{\omega}_k, \bar{\bm{d}}_k\}$ is assumed:
\begin{linenomath*}\begin{align}
    \begin{Bmatrix}
        \bar{\omega}_k \\
        \bar{\bm{d}}_k
    \end{Bmatrix}
    =
    \begin{Bmatrix}
        \hat{\omega}_k \\
        \hat{\bm{d}}_k
    \end{Bmatrix}
    +
    \begin{Bmatrix}
        \varepsilon_{\omega,k} \\
        \bm{\varepsilon}_{\mathrm{d},k}
    \end{Bmatrix}
\end{align}\end{linenomath*}
where $\{ \hat{\omega}_k, {\hat{\bm{d}}_k}^\top \}^\top = \bm{f}(\bm{\theta}_\mathrm{M},\bm{\theta}_\mathrm{K}) \in \mathbb{R}^{N_\mathrm{obs} + 1}$ is a vector of simulated modal properties obtained by solving the eigenvalue problem,
$
    \left( \hat{\omega}_k^2 \mathbf{M}(\bm{\theta}_\mathrm{M}) - \mathbf{K}(\bm{\theta}_\mathrm{K}) \right)
    \hat{\bm{d}}_k = \mathbf{0}
$.
$\{ \varepsilon_{\omega,k}, \bm{\varepsilon}_{\mathrm{d},k}^\top \}^\top \in \mathbb{R}^{N_\mathrm{obs} + 1}$ is a residual vector to represent modeling and measurement error that follows a multivariate Gaussian distribution,
\begin{linenomath*}\begin{align}
    p \left( \begin{Bmatrix} \varepsilon_{\omega,k} \\ \bm{\varepsilon}_{\mathrm{d},k} \end{Bmatrix} \right)
    = \mathcal{N}
    \left(
    \begin{Bmatrix} \varepsilon_{\omega,k} \\ \bm{\varepsilon}_{\mathrm{d},k} \end{Bmatrix}
    ~ \middle | ~
    \begin{Bmatrix} 0 \\ \bm{0} \end{Bmatrix},
    \begin{bmatrix} \sigma^2_\omega & \bm{0} \\ \bm{0} & \mathbf{\Sigma}_\mathrm{dd} \end{bmatrix}
    \right)
\end{align}\end{linenomath*}
In the remaining parts of the paper, we assume an isotropic Gaussian distribution for $\bm{\varepsilon}_{\mathrm{d},k}$ for simplicity, i.e., $\mathbf{\Sigma}_\mathrm{dd} = \sigma_\mathrm{d}^2 \mathbf{I}_{N_\mathrm{obs}}$, where $\mathbf{I}_{N_\mathrm{obs}} \in \mathbb{R}^{N_\mathrm{obs} \times N_\mathrm{obs}}$ denotes an identity matrix.
The \textit{modal-analysis-based likelihood} is thus described as
\begin{linenomath*}\begin{align}
    p(\bar{\omega}_k, \bar{\bm{d}}_k \mid \bm{\theta}_\mathrm{M},\bm{\theta}_\mathrm{K}, \sigma_\omega, \sigma_\mathrm{d})
    = c_1 \exp \left(
    - \frac{(\bar{\omega}_k - \hat{\omega}_k(\bm{\theta}_\mathrm{M},\bm{\theta}_\mathrm{K}))^2}{2 \sigma_\omega^2}
    - \frac{\| \bar{\bm{d}}_k - \hat{\bm{d}}_k(\bm{\theta}_\mathrm{M},\bm{\theta}_\mathrm{K}) \|_2^2}{2 \sigma_\mathrm{d}^2}
    \right)
    \label{eq:modal}
\end{align}\end{linenomath*}
where $c_1$ denotes a normalizing constant.
}

\textcolor{black}{
For the observed MSR $\bar{\bm{r}}_k$, the following generative process is assumed.
\begin{linenomath*}\begin{align}
    \bar{\bm{r}}_k = \hat{\bm{r}}_k + \bm{\varepsilon}_{\mathrm{r},k}
\end{align}\end{linenomath*}
where $\hat{\bm{r}}_k = \bm{g} (\bm{\theta}_\mathrm{K}, \hat{\bm{d}}_k) \in \mathbb{R}^{M}$ is an MSR vector simulated by static analysis as described in Eq.~(\ref{eq:stressres}) given the corresponding MD vector $\hat{\bm{d}}_k$ and the stiffness parameters $\bm{\theta}_\mathrm{K}$.
\begin{linenomath*}\begin{align}
    \hat{\bm{r}}_k = \mathbf{B} \mathbf{K}_\mathrm{m}(\bm{\theta}_\mathrm{K}) \mathbf{H}^\top \mathbf{\Gamma}(\bm{\theta}_\mathrm{K}) \hat{\bm{d}}_k
    \label{eq:static_anal}
\end{align}\end{linenomath*}
where $\mathbf{B} \in \mathbb{R}^{M_\mathrm{obs} \times M}$ represents an observation matrix with entities of zero or unity and $\mathbf{\Gamma} \in \mathbb{R}^{N \times N_\mathrm{obs}}$ is a matrix that transforms a ``reduced'' MD vector for the measured DOFs to a complete MD vector for the full DOFs (refer to Appendix I for details).
$\bm{\varepsilon}_{\mathrm{r},k} \in \mathbb{R}^{M_\mathrm{obs}}$ is a residual vector that follows
\begin{linenomath*}\begin{align}
    p(\bm{\varepsilon}_{\mathrm{r},k}) = \mathcal{N}(\bm{\varepsilon}_{\mathrm{r},k} \mid \mathbf{0}, \mathbf{\Sigma}_\mathrm{rr})
\end{align}\end{linenomath*}
$\mathbf{\Sigma}_\mathrm{rr} = \sigma_\mathrm{r}^2 \mathbf{I}_{M_\mathrm{obs}} \in \mathbb{R}^{M_\mathrm{obs} \times M_\mathrm{obs}}$ is assumed in the remaining parts of the paper.
The \textit{static-analysis-based likelihood} is expressed as
\begin{linenomath*}\begin{align}
    p(\bar{\bm{r}}_k \mid \bm{\theta}_\mathrm{K}, \hat{\bm{d}}_k, \sigma_\mathrm{r}) =
    c_2 \exp \left( -\frac{\| \bar{\bm{r}}_k - \hat{\bm{r}}_k (\bm{\theta}_\mathrm{K}, \hat{\bm{d}}_k) \|_2^2}{2 \sigma_\mathrm{r}^2} \right)
    \label{eq:static}
\end{align}\end{linenomath*}
where $c_2$ is a normalizing constant.
}

\textcolor{black}{
From Eqs.~(\ref{eq:modal}), (\ref{eq:static}), and referring to Bayes' Theorem, the posterior probability density function (PDF) of structural parameters $\bm{\theta}_\mathrm{M}, \bm{\theta}_\mathrm{K}$ and statistical parameters $\sigma_\omega, \sigma_\mathrm{d}, \sigma_\mathrm{r}$ is obtained by
\begin{linenomath*}\begin{align}
    ~ & ~
    p(\bm{\theta}_\mathrm{M}, \bm{\theta}_\mathrm{K}, \sigma_\omega, \sigma_\mathrm{d}, \sigma_\mathrm{r} \mid \bar{\bm{\omega}}, \bar{\bm{d}}, \bar{\bm{r}})
    \nonumber
    \\
    \propto & ~
    p(\bar{\bm{\omega}}, \bar{\bm{d}}, \bar{\bm{r}} \mid \bm{\theta}_\mathrm{M}, \bm{\theta}_\mathrm{K}, \sigma_\omega, \sigma_\mathrm{d}, \sigma_\mathrm{r}) ~
    p(\bm{\theta}_\mathrm{M}, \bm{\theta}_\mathrm{K}, \sigma_\omega, \sigma_\mathrm{d}, \sigma_\mathrm{r})
    \nonumber \\
    = & ~
    \prod_{k=1}^{K}
    p(\bar{\omega}_k, \bar{\bm{d}}_k \mid \bm{\theta}_\mathrm{M}, \bm{\theta}_\mathrm{K}, \sigma_\omega, \sigma_\mathrm{d}) ~
    p(\bar{\bm{r}}_k \mid \bm{\theta}_\mathrm{K}, \hat{\bm{d}}_k, \sigma_\mathrm{r}) ~
    p(\bm{\theta}_\mathrm{M}, \bm{\theta}_\mathrm{K}, \sigma_\omega, \sigma_\mathrm{d}, \sigma_\mathrm{r})
    \nonumber \\
    \propto & ~
    \exp \left (
    - \frac{(\bar{\omega}_k - \hat{\omega}_k)^2}{2 \sigma_\omega^2}
    - \frac{\| \bar{\bm{d}}_k - \hat{\bm{d}}_k \|_2^2}{2 \sigma_\mathrm{d}^2}
    - \frac{\| \bar{\bm{r}}_k - \hat{\bm{r}}_k \|_2^2}{2 \sigma_\mathrm{r}^2}
    \right ) ~
    p(\bm{\theta}_\mathrm{M}, \bm{\theta}_\mathrm{K}, \sigma_\omega, \sigma_\mathrm{d}, \sigma_\mathrm{r})
    \label{eq:posterior}
\end{align}\end{linenomath*}
The equality in Eq.~(\ref{eq:posterior}) holds under the assumption of statistical independence among different vibration modes.
For the prior PDF, $p(\bm{\theta}_\mathrm{M}, \bm{\theta}_\mathrm{K}, \sigma_\omega, \sigma_\mathrm{d}, \sigma_\mathrm{r})$, we place \textit{weakly informative priors} \citep{gelmanA2013} to regularize the parameters while preventing a strong bias.
The implementation details are described in the NUMERICAL EXAMPLE and EXPERIMENTAL VALIDATION sections.
}

\subsection{MCMC Sampling}

\textcolor{black}{
In this study, the posterior in Eq.~(\ref{eq:posterior}) was evaluated numerically by using Markov chain Monte Carlo (MCMC) simulations;
specifically, the No-U-Turn Sampler (NUTS) algorithm \citep{hoffmanMD2014}, an improved class of the Hamiltonian Monte Carlo (HMC) algorithm \citep{nealRM2011}, was adopted.
Based on the analogy of Hamiltonian dynamics, HMC introduces a parameter vector $\bm{\theta}$ as well as an auxiliary vector $\bm{\varphi}$ that has the same dimensions as $\bm{\theta}$.
$\bm{\theta}, \bm{\varphi}$ corresponds to the position and momentum of particles in Hamiltonian dynamics, respectively.
HMC samples $\{\bm{\theta}, \bm{\varphi}\} \sim p(\bm{\theta}\mid \mathcal{D})p(\bm{\varphi})$ by simulating Hamiltonian dynamics, where the logarithm of the joint posterior, $\log p(\bm{\theta}, \bm{\varphi}\mid \mathcal{D}) = \log p(\bm{\theta}\mid \mathcal{D}) + \log p(\bm{\varphi})$, are considered as ``Hamiltonian'' $H$, i.e., the sum of potential and kinetic energy.
The algorithm is summarized as follows.
}

\textcolor{black}{
\textit{HMC Algorithm} \citep{nealRM2011}.
Let $\bm{\theta}^0$ be the initial value of $\bm{\theta}$.
Let $\epsilon$ and $L$ be the step size and the number of steps in the leapfrog algorithm.
Define the prior PDF of $\bm{\varphi}$ as $p(\bm{\varphi}) = \mathcal{N}(\bm{\varphi} \mid \mathbf{0}, \mathbf{\Sigma})$.
For $t = 1,..., T$, repeat the following procedure:
\begin{enumerate}
    \item Initialize the momentum vector:
    $\bm{\varphi}^0 \sim p(\bm{\varphi})$
    \item Set the candidate vectors: $\bm{\theta}^\ast \leftarrow \bm{\theta}^{t-1}$; $\bm{\varphi}^\ast \leftarrow \bm{\varphi}^0$
    \item The leapfrog algorithm: repeat the following procedure $L$ times.
    \begin{enumerate}
        \item Update the momentum vector
        \begin{linenomath*}\begin{align}
            \bm{\varphi}^\ast \leftarrow \bm{\varphi}^\ast + {\displaystyle \frac{1}{2} \epsilon \frac{\mathrm{d} \log p(\bm{\theta}^\ast \mid \mathcal{D})}{\mathrm{d} \bm{\theta}^\ast}}
        \end{align}\end{linenomath*}
        \item Update the position vector
        \begin{linenomath*}\begin{align}
            \bm{\theta}^\ast \leftarrow \bm{\theta}^\ast + \epsilon \mathbf{\Sigma}^{-1} \bm{\varphi}^\ast
        \end{align}\end{linenomath*}
        \item Update the momentum vector
        \begin{linenomath*}\begin{align}
            \bm{\varphi}^\ast \leftarrow \bm{\varphi}^\ast + {\displaystyle \frac{1}{2} \epsilon \frac{\mathrm{d} \log p(\bm{\theta}^\ast \mid \mathcal{D})}{\mathrm{d} \bm{\theta}^\ast}}
        \end{align}\end{linenomath*}
    \end{enumerate}
    \item Calculate the following density ratio,
    \begin{linenomath*}\begin{align}
        r = \frac{p(\bm{\theta}^\ast \mid \mathcal{D})p(\bm{\varphi}^\ast)}{p(\bm{\theta}^{t-1} \mid \mathcal{D})p(\bm{\varphi}^0)}
        \label{eq:ratio}
    \end{align}\end{linenomath*}
    \item Accept or reject the candidate sample
    \begin{linenomath*}\begin{align}
        \label{eq:accept}
        \bm{\theta}^t =
        \left \{
        \begin{array}{ll}
            \bm{\theta}^\ast  & \text{with the probability of } \mathrm{min}(r, 1)\\
            \bm{\theta}^{t-1} & \text{otherwise}
        \end{array}
        \right .
    \end{align}\end{linenomath*}
\end{enumerate}
Thus we obtain the posterior samples $\bm{\theta}_t \sim p(\bm{\theta} \mid \mathcal{D}), ~ t = T_0 + 1, ..., T$, where the burn-in period is defined as $[0, T_0]$.
}

\textcolor{black}{
Hamiltonian dynamics has the properties of time-reversibility and volume-preservation, thereby ensuring ``detailed balance''---a sufficient condition for the convergence of simulated samples to a stationary distribution \citep{nealRM2011}.
Furthermore, when the Hamiltonian is time-invariant, indicating that energy conservation is satisfied, the acceptance probability in Eq.~(\ref{eq:ratio}) equals one. Although energy conservation is not necessarily satisfied due to approximation errors in the implementation, HMC achieves a remarkable sampling efficiency.
}

\textcolor{black}{
Despite the superior performance of HMC, the hyperparameters $\epsilon, L$ in the leapfrog algorithm should be carefully determined because of their significance on sampling efficiency.
The NUTS algorithm alleviates the burden of manual tuning, where $\epsilon, L$ are not fixed and automatically adjusted in each iteration $t$.
The readers are referred to \citet{hoffmanMD2014} for details.
}

\subsection{Discussions}

\textcolor{black}{
Many existing methods in the Bayesian FE model updating literature are based on the modal-analysis-based likelihood as expressed by Eq.~(\ref{eq:modal}).
Given that an infinite number of combinations of $\{\mathbf{M}, \mathbf{K}\}$ satisfy the eigenvalue problem in Eq.~(\ref{eq:eigen}) \citep{beckJL2002},
the evaluation of stiffness parameters (and structural damages) requires a prior assumption on mass parameters $\bm{\theta}_\mathrm{M}$ to constrain the feasible space of stiffness parameters $\bm{\theta}_\mathrm{K}$.
}

\textcolor{black}{
The proposed stress-resultant-based approach exploits the relationship between nodal displacement $\bm{d}$ and stress-resultant $\bm{r}$ in the modal space, by applying system identification techniques jointly to acceleration and dynamic strain measurements.
This relationship enables the MSR vector to be normalized by the MD vector, providing a sensitive measure for elemental stiffness referred to as ``local stiffness'' in \citet{iyamaJ2021} and \citet{iyamaJ2023}.
For example, in the case of a single-bay, single-story 2D frame subject to lateral excitation, such normalized MSR vector $\bar{\bm{r}}_k$ characterizes the stress-resultant distribution in a frame when a unit displacement is applied at the top of the frame.
The decrease in the MSR thus indicates the decrease in the contribution of the local element to the (global) restoring force of the entire frame, suggesting the existence of localized structural damage.
}

\textcolor{black}{
Eq.~(\ref{eq:static_anal}) suggests that the relationship between MD and MSR can be established via static analysis given the stiffness parameters $\bm{\theta}_\mathrm{K}$.
Conversely, $\bm{\theta}_\mathrm{K}$ can be identified by solving the inverse problem according to a given MD and MSR set; notably this procedure is independent of any mass assumptions, providing a basis for the static-analysis-based likelihood in Eq.~(\ref{eq:static}).
The proposed formulation combines this static-analysis-based likelihood with the conventional modal-analysis-based likelihood.
In the exponential term in Eq.~(\ref{eq:posterior}), the static-analysis-based likelihood acts as a form of ``regularization'' term to constrain the space of $\bm{\theta}_\mathrm{K}$ within the modal-analysis-based likelihood, allowing for simultaneous identification of $\bm{\theta}_\mathrm{M}$ and $\bm{\theta}_\mathrm{K}$.
}

\section{Numerical Example}

\subsection{Structural Model}

\begin{figure}[t]
    \centering
    \includegraphics[height=80truemm]{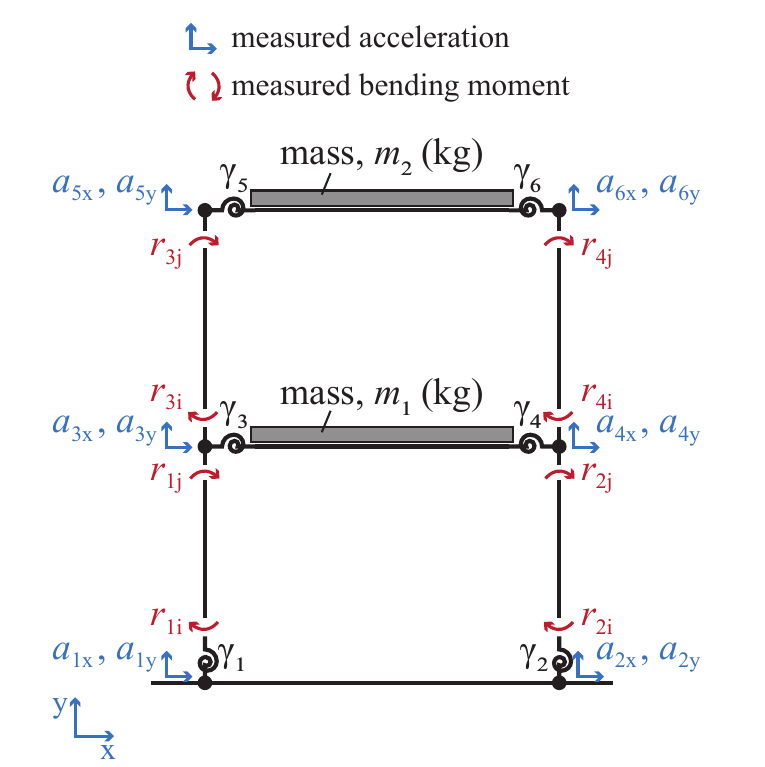}
    \caption{One-bay two-story planar frame for numerical example.}
    \label{fig:num-example-sec}
\end{figure}

In this section, a numerical example is presented using a two-story, one-bay, moment-resisting planar frame, as depicted in Fig.~1.
The frame, assumed to be a steel structure, possesses specific material and mechanical properties: Young's modulus $E = 2.05 \times 10^5~(\text{N/mm}^2)$; the cross-sectional area of the columns $A_\mathrm{c} = 6.67 \times 10^3~(\text{mm}^2)$; the second moment of the area of the columns $I_\mathrm{c} = 3.99 \times 10^7~(\text{mm}^4)$; the cross-sectional area of the beams $A_\mathrm{b} = 4.68 \times 10^3~(\text{mm}^2)$; and the second moment of the area of the beams $I_\mathrm{b} = 7.21 \times 10^3~(\text{mm}^4)$.
These properties are assumed to be known in advance.
The shear deformation of the elements is neglected, i.e., the shear modulus $G = \infty$.
In addition to the intrinsic weights of the elements with a density of $\rho = 7.85 \times 10^3 ~ (\text{kg/m}^3)$, an additional mass $m_j$ was allocated to each floor $j ~ (= 1, 2)$.
Half of the mass was lumped at both ends of the floor beam, i.e., a lumped mass matrix was considered.
A stiffness-proportional damping matrix was employed with a damping ratio of $0.02$.
To model the structural damage to the beam ends and column bases, rotational springs with stiffnesses of $k_i ~ (\text{kNm}/\text{rad}) ~ (i = 1, ..., 6)$ were considered as illustrated in Fig.~1.
To simplify the implementation, these rotational stiffness values were converted into \textit{fixity factors} \citep{dhillonBS1990}:
\begin{linenomath*}\begin{align}
    \gamma_i = \left( 1 + \frac{3 EI_i/L}{k_i} \right)^{-1}  \label{eq:fixity}
\end{align}\end{linenomath*}
where $I_i = I_\mathrm{b} ~ (i = 1, 2)$ or $I_i = I_\mathrm{c} ~ (i = 3, ..., 6)$.
$\gamma_i \in [0, 1]$ represents a perfectly rigid joint when $\gamma_i = 1$ and a pinned joint when $\gamma_i = 0$.

The parameters to be estimated include the structural parameters $\bm{\theta}_\mathrm{M} = [m_1, m_2]^\top$ and $\bm{\theta}_\mathrm{K} = [\gamma_1,...,\gamma_6]^\top$, with no prior assumptions, and the statistical parameters $\sigma_\mathrm{d}, ~\sigma_\mathrm{r}$, and $\sigma_\omega$.
The target values were set as $\gamma_1 = 0.3,~\gamma_2 = 0.5,~\gamma_3 = 0.7,~\gamma_4 = 0.8,~\gamma_5 = 1.0,~\gamma_6 = 1.0,~m_1 = 2000 ~ (\mathrm{kg}), m_2 = 1000 ~ (\mathrm{kg})$.
Herein, we assumed severe damage to both column bases and moderate damage to both ends of the second-floor beam.

\subsection{Synthetic Data}

\begin{figure}[t]
    \centering
    \includegraphics[width=80truemm]{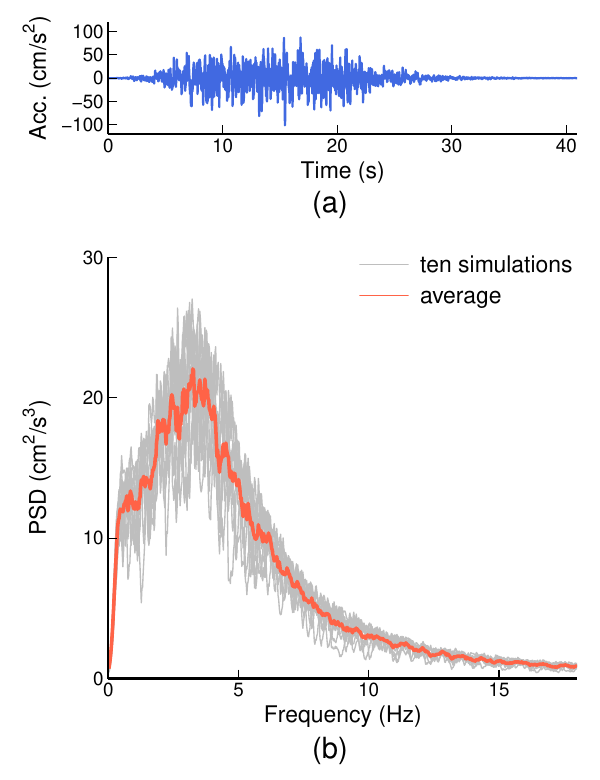}
    \caption{Samples of simulated ground motion: (a) time history of one simulation, and (b) power spectral density (PSD) of 10 simulations smoothed by a Parzen window with a bandwidth of 0.5 Hz.}
    \label{fig:iac}
\end{figure}

Synthetic data comprising nodal acceleration and elemental stress resultant (specifically, bending moment responses) were generated through elastic time-history analyses for the target frame.
As depicted in Fig.~1, twelve nodal acceleration components, including six horizontal components $a_\mathrm{1x},...,a_\mathrm{6x}$ and six vertical components $a_\mathrm{1y},...,a_\mathrm{6y}$ (with the constraints of $a_\mathrm{1x} = a_\mathrm{2x}$ and $a_\mathrm{1y} = a_\mathrm{2y}$), and eight bending moment components $r_\mathrm{1i}, r_\mathrm{1j}, ..., r_\mathrm{4j}$ were used as measurements, assuming that bending moment responses were obtained by dynamic strain measurements with known mechanical and material properties ($EI_\mathrm{c}$).
Time history analyses focused solely on a lateral excitation, modeled as a zero-mean, non-stationary random process.
Initially, a stationary process with a duration of 40.96s and a sampling rate of 100 Hz, which adheres to a Kanai--Tajimi power spectral density (PSD) with a dominant angular frequency of $\omega_g = 8 \pi$ (rad/s), a bandwidth parameter of $\zeta = 0.6$, and an intensity parameter of $\Phi_0 = 5.0$, was synthesized using the spectral representation method \citep{shinozukaM1991}.
The envelope function proposed by \citet{aminM1968}, with a stationary part within $[8.20\text{s}, 20.48\text{s}]$, was applied to simulate non-stationarity, and a cosine-tapered high-pass filter was used to reduce the frequency content below $0.5$ Hz.
\textcolor{black}{
Fig.~2(a) illustrates a sample of the input ground motion time histories;
Fig.~2(b) depicts the PSD of ground motions in 10 different simulations and their averaged spectrum.
}
The observation noise is represented as a Gaussian white noise and added to both excitation and response.
To investigate the impact of noise levels on model parameter estimation, different scenarios of observation noise were examined.
Two cases of the s.d.~of the observation noise for acceleration time histories (including both excitation and responses) were considered: $0.01 \sigma_\mathrm{5x}$ and $0.1 \sigma_\mathrm{5x}$, where $\sigma_\mathrm{5x}$ denotes the root mean square (RMS) value of the acceleration response $a_\mathrm{5x}$.
Similarly, for the bending moment responses, two cases of s.d.~were assessed as $0.01 \sigma_{\mathrm{1i}}$ and $0.1 \sigma_{\mathrm{1i}}$, with $\sigma_{\mathrm{1i}}$ representing the RMS value of the bending moment response $r_\mathrm{1i}$.
These variations led to four distinct cases, Cases 1--4, as outlined in Table~\ref{tab:noise}.

\begin{table}
    \centering
    \fontsize{10truept}{11truept}\selectfont
    \caption{S.D.~of imposed observation noises for all cases. $\sigma_{5x}$ and $\sigma_{1i}$ denote the RMS values of the acceleration response $a_\mathrm{5x}$ and bending moment response $r_\mathrm{1i}$.}\label{tab:noise}
    \begin{tabular*}{\textwidth}{@{\extracolsep{\fill}}rrr}
        \toprule
        Case & For Acc. & For Bending Moment \\
        \midrule
        1 & $0.01 \times \sigma_\mathrm{5x}$ & $0.01 \times \sigma_\mathrm{1i}$ \\
        2 & $0.10 \times \sigma_\mathrm{5x}$ & $0.01 \times \sigma_\mathrm{1i}$ \\
        3 & $0.01 \times \sigma_\mathrm{5x}$ & $0.10 \times \sigma_\mathrm{1i}$ \\
        4 & $0.10 \times \sigma_\mathrm{5x}$ & $0.10 \times \sigma_\mathrm{1i}$ \\
        \bottomrule
    \end{tabular*}
\end{table}

\subsection{Modal Identification}

System identification was conducted using the MOESP algorithm \citep{verhaegenM1992a}, a subspace state-space system identification (4SID) method, on the input--output data.
These data contained two components of ground acceleration ($a_\mathrm{1x} (= a_\mathrm{2x})$ and $a_\mathrm{1y} (= a_\mathrm{2y})$) as inputs, in addition to the other eight acceleration components and all eight bending moment components as outputs.
As the critical hyperparameters in the MOESP implementation, the number of block rows in a block Hankel matrix and the system order were configured to 30 and 10, respectively.
A successful MOESP execution typically produces conjugate pairs of complex modes. Thus, an implementation with a system order of 10 practically yielded five distinct vibration modes, with the first two modes (denoted as Modes 1 and 2) being used as ``observation'' in the following Bayesian inference.

Consequently, a dataset of modal properties, including natural angular frequencies $\bar{\omega}_k$, mode shapes in terms of acceleration, termed \textit{modal acceleration (MA)}, $\bar{\bm{a}}^\mathrm{c}_k$, and MSR $\bar{\bm{r}}^\mathrm{c}_k$, was obtained for $k = 1, 2$.
The superscript $\mathrm{c}$ indicates complex mode shapes.
\textcolor{black}{
Although MSR may include the axial force and bending moment, this example considers the mode shapes only in terms of bending moment, which in the following are specifically termed \textit{modal bending moment (MBM)}.
}
The MA vector $\bar{\bm{a}}^\mathrm{c}_k$ was then transformed into the MD vector $\bar{\bm{d}}^\mathrm{c}_k$ as follows \citep{yaoyamaT2023}:
\begin{linenomath*}\begin{align}
    \bar{\bm{d}}^\mathrm{c}_k = - \frac{\bar{\bm{a}}^\mathrm{c}_k}{\bar{\omega}_k^2} \label{eq:matomd}
\end{align}\end{linenomath*}
A real eigenvalue problem is considered in the likelihood evaluation; hence, the complex mode shapes $\bar{\bm{d}}^\mathrm{c}_k$ and $\bar{\bm{r}}^\mathrm{c}_k$ were transformed into real mode shapes as follows:
\begin{linenomath*}\begin{align}
    \bar{\bm{d}}_k &= \mathrm{Re} \left( \bar{\bm{d}}^\mathrm{c}_k ~ \mathrm{e}^{-\mathrm{i}\theta_{k,\mathrm{ref}}^\mathrm{d}} \right) \\
    \bar{\bm{r}}_k &= \mathrm{Re} \left( \bar{\bm{r}}^\mathrm{c}_k ~ \mathrm{e}^{-\mathrm{i}\theta_{k,\mathrm{ref}}^\mathrm{r}} \right)
\end{align}\end{linenomath*}
where $\mathrm{i} = \sqrt{-1}$, and $\mathrm{Re}(\cdot)$ denotes the real part of a complex value.
$\theta_{k,\mathrm{ref}}^\mathrm{d}$ and $\theta_{k,\mathrm{ref}}^\mathrm{r}$ represent the phase of mode shapes at reference components $\bar{d}_{k,\mathrm{ref}} \in \bar{\bm{d}}_k$ and $\bar{r}_{k,\mathrm{ref}} \in \bar{\bm{r}}_k$.
In this example, $d_\mathrm{3x}$ (corresponding to $a_\mathrm{3x}$) and $r_\mathrm{1i}$ were used as reference components for both modes $k = 1,2$.
This process deals with an unexpected phase difference between MD and MBM, which might be attributed to the imposed observation noise or the relatively low accuracy of mode shape estimation in system identification.
Notably, the MD and MBM vectors $\{\bar{\bm{d}}_k,\bar{\bm{r}}_k\}$ were normalized by the same factor, such that the Euclidean norm of the MD vector $\|\bar{\bm{d}}_k\|_2 = 1$ and the relationship between $\bar{\bm{d}}_k$ and $\bar{\bm{r}}_k$ was preserved for $\bar{\bm{r}}_k$ to represent the internal force corresponding to external deformation $\bar{\bm{d}}_k$.

\begin{figure}[t]
    \centering
    \includegraphics[width=\textwidth]{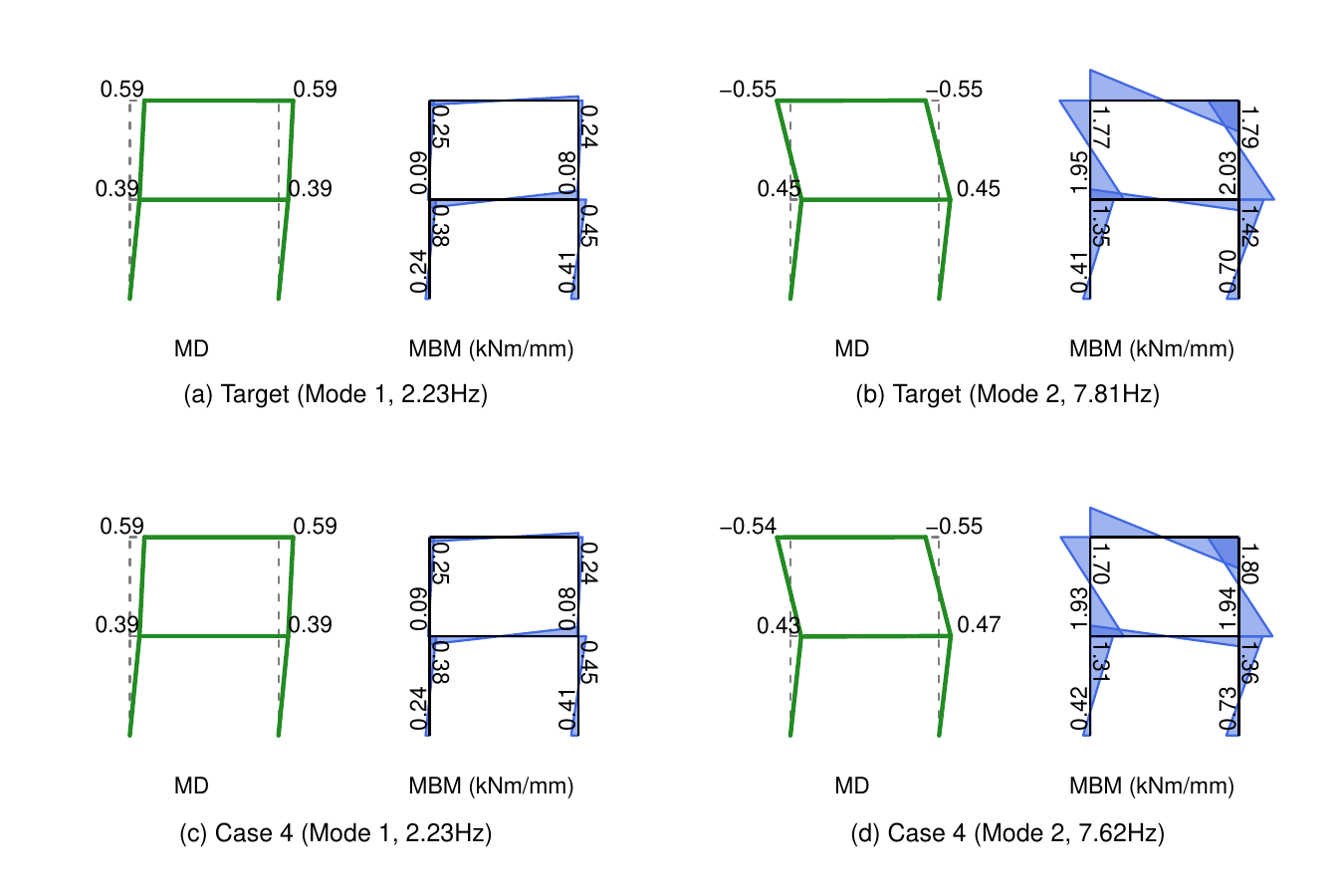}
    \caption{Modal displacement (MD) and modal bending moment (MBM) for Modes 1 and 2: (a)--(b) target mode shapes; (c)--(d) identified mode shapes in a simulation for Case 4.}
    \label{fig:modes}
\end{figure}

Reiterating the aforementioned process yielded ten sets of synthesized data, each with distinct excitations and observation noises, to examine the impact of different realizations on Bayesian inference. 
Fig.~3 illustrates the mode shapes (MD and MBM) of the target structure alongside those identified in a simulation of Case 4 (the scenario with the highest level of contamination).
In both modes, the MBM values at the base of the left column are remarkably lower than those at the base of the right column, indicating a reduced fixity factor at the left column base ($\gamma_1 < \gamma_2$). For Mode 1, the identified modal properties, including the natural frequency, MD, and MBM, are in close agreement with the target properties.
However, in Mode 2, the identified natural frequency is 2--3\% lower than the target, with minor deviations in the identified MD and MBM values. 
This pattern persists across the other nine simulations, although not depicted due to spatial constraints.
The implications of this trend on parameter estimation are further explored in subsequent subsections.

\subsection{Bayesian Inference}

For the stiffness parameters $\gamma_i ~ (i = 1, ..., 6)$, a uniform prior with the same range as the domain of $\gamma_i$, i.e., $\mathcal{U}(0,1)$ was assigned.
Similarly, for the mass parameters $m_1, m_2$ (kg), a uniform prior with a sufficiently broad range~$\mathcal{U}(0, 5 \times 10^4)$ was considered.
For better convergence, even in situations of data scarcity, weakly informative priors \citep{gelmanA2013} were used for $\sigma_\mathrm{d}, \sigma_\mathrm{r}, \sigma_\omega$.
\begin{linenomath*}\begin{align}
    \sigma_\mathrm{d} & \sim \mathcal{N}^+ (0, s_\mathrm{d}^2) \label{eq:weakinfo-d} \\
    \sigma_\mathrm{r} & \sim \mathcal{N}^+ (0, s_\mathrm{r}^2) \label{eq:weakinfo-r} \\
    \sigma_\mathrm{\omega} & \sim \mathcal{N}^+ (0, s_\mathrm{\omega}^2) \label{eq:weakinfo-w}
\end{align}\end{linenomath*}
where $\mathcal{N}^+(.,.)$ represents a truncated normal distribution defined over a random variable greater than or equal to zero.
For example, the prior PDF in Eq.~(\ref{eq:weakinfo-r}) indicates that $\sigma_\mathrm{r}$ is unlikely to exceed $s_\mathrm{r}$ but allows exceedance with a small probability.
The hyperparameters were set as $s_\mathrm{r} = 50~(\text{kNm/m}), ~ s_\mathrm{d} = 0.05, ~ s_\omega = 0.4 \pi~(\text{rad/s})$.

The NUTS algorithm for MCMC simulation was implemented using the Stan software \citep{stan}.
Four chains with different initial values were set, and 2000 samples including a burn-in period of 1000 samples were generated for each chain.
For faster convergence, the initial values for $m_1,m_2$ were randomly sampled from the range $[0, 2.5 \times 10^4]$, which is half the domain of the prior PDFs (although the upper bound is considerably larger than the targets).
The initial values of the additional parameters were randomly sampled from the entire domain of the prior PDFs.
The convergence of all the parameters was confirmed by reviewing that a convergence metric, called the potential scale reduction factor $\widehat{R}$, was less than 1.1 \citep{gelmanA2013}.

\begin{figure}[!t]
    \centering
    \includegraphics[width=135truemm]{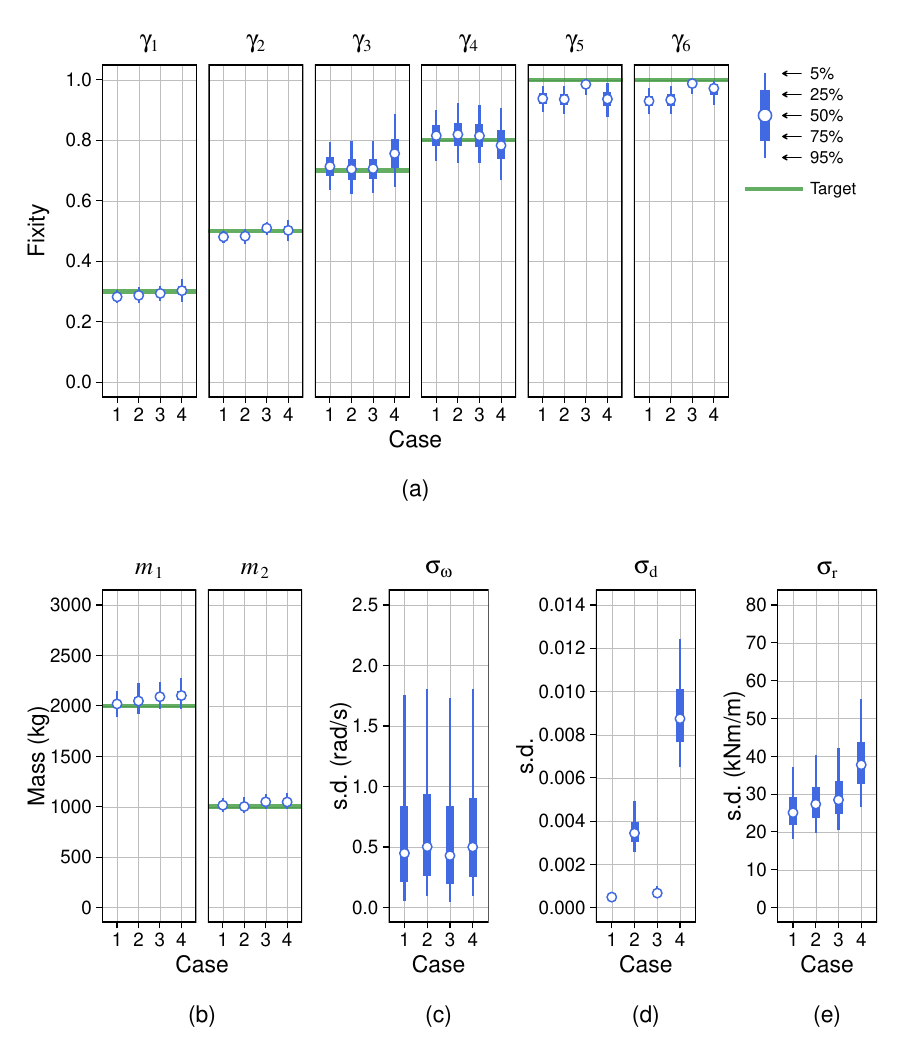}
    \caption{Distributions of the posterior samples in a simulation for Cases 1--4: (a) stiffness parameters; (b) mass parameters; (c) $\sigma_\omega$; (d) $\sigma_\mathrm{d}$; (e) $\sigma_\mathrm{r}$.}
    \label{fig:num-posterior}
\end{figure}

\begin{table}[!t]
    \fontsize{10truept}{11truept}\selectfont
    \centering
    \caption{Mean and CoV of the posterior mean values in all simulations.}\label{tab:meancov}
    \begin{tabular*}{\textwidth}{@{\extracolsep{\fill}}lrrrr}
        \toprule
        Parameter & Target & Case & Mean of posterior mean & CoV of posterior mean \\
        \midrule
        $\gamma_1$ & 0.3  & 1 & 0.284 & 0.007 \\  
        ~          & ~    & 2 & 0.283 & 0.024 \\  
        ~          & ~    & 3 & 0.304 & 0.057 \\  
        ~          & ~    & 4 & 0.299 & 0.049 \\  
        $\gamma_2$ & 0.5  & 1 & 0.479 & 0.003 \\  
        ~          & ~    & 2 & 0.477 & 0.018 \\  
        ~          & ~    & 3 & 0.503 & 0.041 \\  
        ~          & ~    & 4 & 0.490 & 0.032 \\  
        $\gamma_3$ & 0.7  & 1 & 0.710 & 0.008 \\  
        ~          & ~    & 2 & 0.720 & 0.015 \\  
        ~          & ~    & 3 & 0.736 & 0.048 \\  
        ~          & ~    & 4 & 0.720 & 0.060 \\  
        $\gamma_4$ & 0.8  & 1 & 0.818 & 0.007 \\  
        ~          & ~    & 2 & 0.820 & 0.014 \\  
        ~          & ~    & 3 & 0.820 & 0.060 \\  
        ~          & ~    & 4 & 0.821 & 0.052 \\  
        $\gamma_5$ & 1.0  & 1 & 0.934 & 0.004 \\  
        ~          & ~    & 2 & 0.924 & 0.020 \\  
        ~          & ~    & 3 & 0.971 & 0.017 \\  
        ~          & ~    & 4 & 0.962 & 0.036 \\  
        $\gamma_6$ & 1.0  & 1 & 0.928 & 0.005 \\  
        ~          & ~    & 2 & 0.925 & 0.020 \\  
        ~          & ~    & 3 & 0.981 & 0.008 \\  
        ~          & ~    & 4 & 0.974 & 0.020 \\
        $m_1$ (kg) & 2000 & 1 & 2026  & 0.003 \\
        ~          & ~    & 2 & 2038  & 0.015 \\
        ~          & ~    & 3 & 2099  & 0.017 \\  
        ~          & ~    & 4 & 2102  & 0.014 \\
        $m_2$ (kg) & 1000 & 1 & 1014  & 0.003 \\
        ~          & ~    & 2 & 1011  & 0.011 \\
        ~          & ~    & 3 & 1054  & 0.017 \\
        ~          & ~    & 4 & 1043  & 0.025 \\
        \bottomrule
    \end{tabular*}
\end{table}

\begin{figure}[t]
    \centering
    \includegraphics[width=150truemm]{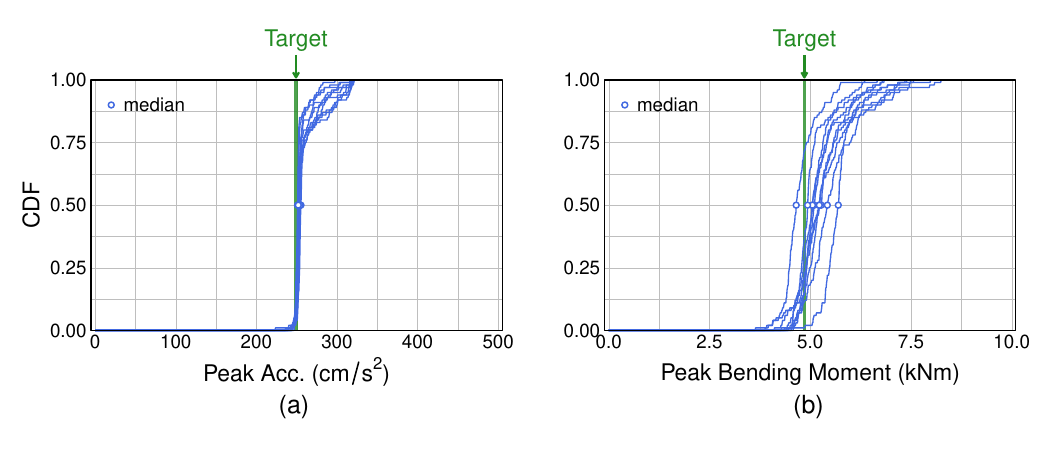}
    \caption{Predictive cumulative density functions (CDFs) of response to an ``unobserved'' excitation for 10 different simulations: (a) CDF of peak acceleration of $a_\mathrm{5x}$; (b) CDF of peak bending moment of $r_\mathrm{1i}$.}
    \label{fig:num-pred}
\end{figure}

Fig.~4 presents the posterior sample distributions for all parameters across all cases.
As depicted in Fig.~4(a), the medians of the posterior PDFs for $\gamma_1, ..., \gamma_4$ are well aligned with the target values, and the 5th to 95th percentile ranges (90\% credible intervals) encompass these targets.
However, particularly in Cases 1 and 2, the medians for $\gamma_5$ and $\gamma_6$ diverge notably from the targets, with even the 5th percentile values falling below them.
This discrepancy might originate from the numerical challenges in managing the upper limit ($=1$) of the fixity factors (i.e., $k_5 = k_6 = \infty$) and/or the marginally reduced accuracy in MBM estimation, as illustrated in Fig.~3.
Nevertheless, this bias is largely negligible, as \citet{eurocode3} indicates that joints with fixity factors above 0.893 ($k_i/(EI_i/L) > 25$) can be treated as rigid connections in nonbraced steel frame structures.
Across all cases, Fig.~4(b) demonstrates that the target values for the mass parameters $m_1,m_2$ closely approximate the medians and fall within the 90\% credible intervals.
These findings collectively suggest that the proposed method can estimate the mass and stiffness parameters with reduced and quantified uncertainties, thereby improving structural damage localization and assessment.
Fig.~4(c)--(e) display the posterior sample distributions for the statistical parameters.
The posterior PDF of $\sigma_\omega$ remains relatively consistent across different cases, whereas that of $\sigma_\mathrm{d}$ exhibits significant variation, with markedly higher values in Cases 2 and 4 reflecting increased noise levels.
Conversely, the variation in the posterior PDF of $\sigma_\mathrm{r}$ is not pronounced across Cases 1--4, possibly because of the relatively reduced accuracy in MBM assessment, even in Cases 1 and 2 with lower noise levels.

Table~\ref{tab:meancov} presents the mean and coefficient of variation (CoV) of the posterior mean values across all ten simulations.
Notwithstanding a discernible bias in the estimation of rigid connections, the mean of the posterior mean values aligns closely with the targets, with CoV values for most parameters considerably below 0.1.
This indicates the robustness of the parameter estimation against varying input ground motions and observation noise.

To evaluate the prediction performance of the proposed method, the predictive PDFs of the quantities of interest (QoI) were calculated for all Case 4 simulations.
The selected QoI were the peak acceleration response at the frame's top, $a_\mathrm{5x}$, and the peak bending moment response at a column base, $r_\mathrm{1i}$, in response to an ``unobserved'' ground motion, a different realization from the PSD used for the synthesized data.
For each simulation, QoI samples were derived through time-history analyses using 100 samples of structural parameters extracted at every 40th interval from the 4000 posterior samples.
Fig.~5 displays the predictive cumulative density functions (CDFs) for (a) the peak acceleration of $a_\mathrm{5x}$ and (b) the peak bending moment of $r_\mathrm{1i}$ across all ten simulations.
In Fig.~5(a), the medians closely correspond with the target, and the steep slopes around the medians indicate a precise prediction of peak acceleration within a narrow uncertainty range.
Conversely, the CDF curves in Fig.~5(b) exhibit more gradual slopes near the median, reflecting greater uncertainties in the predictions.
Moreover, these curves exhibit notable dispersion across different simulations, suggesting that bending moment predictions are more susceptible to variations in ground motions and noise in the observed dataset.
This might be attributed to the higher sensitivity of the bending moment to uncertainties in structural parameters, particularly $\gamma_3, \gamma_4$.
Despite these uncertainties, the target values are generally within the slope ranges, indicating that the proposed approach facilitates reasonable QoI prediction.

\section{Experimental Validation}

\subsection{Target Structure and Data}

\begin{figure}[!t]
    \centering
    \includegraphics[width=150truemm]{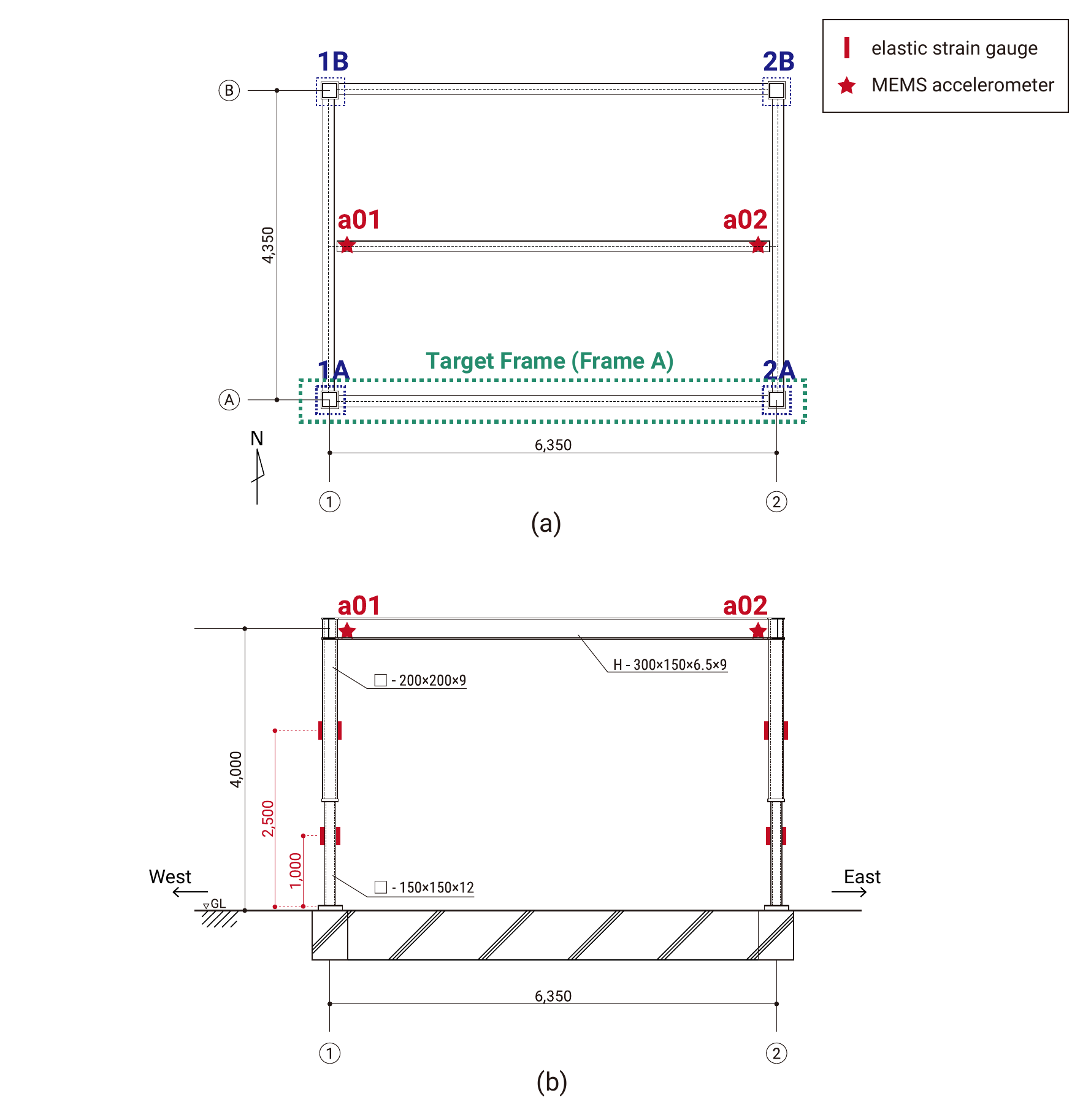}
    \caption{The target structure in the experimental validation: (a) a planar view; (b) a cross-sectional view along A-axis (Frame A).}
    \label{fig:view}
\end{figure}

In this section, an experimental validation utilizing a full-scale moment-resisting steel frame is detailed, as illustrated in Fig.~6 (refer to \citet{yaoyamaT2023} for details of the experimental setup). 
The frame features four columns (Columns 1A, 2A, 1B, and 2B), each with varying box cross-sections in their upper and lower segments. 
The roof is a composite structure comprising H-beams and a reinforced concrete (RC) slab.
The structural damage at the exposed column bases was simulated by loosening the anchor bolts connecting the base plates to the RC footing.
Dynamic strain at two cross-sections, one each in the upper and lower parts, was measured for every column using elastic strain gauges (YEFLAB-5, Tokyo Measuring Lab.)~affixed to the four edges of each cross-section.
Additionally, 3-axis MEMS accelerometers (ADXL335, Analog Devices) were installed at each end of the roof purlin isolated from the main structure.
This setup resulted in 32 strain response components ($= 4~\text{edges} \times 2~\text{cross-sections} \times 4~\text{columns}$) and 4 acceleration response components (horizontal and vertical components from two accelerometers).
The responses were acquired using wireless sensor networks via Raspberry Pi units connected to Wi-Fi.
Both strain and acceleration signals were synchronized through linear interpolation based on the timestamps of each unit, filtering out the frequency content below 1 Hz and above 10 Hz with a cosine-tapered bandpass filter.
The synchronized sampling period was set at 0.046 s. 

The free-vibration responses of the target structure were induced through manual excitation along the EW direction for different damage scenarios (Cases 0--5), as outlined in Table~\ref{tab:cases}. 
Case 0 denotes the original intact state.
In Case 1, a decrease in stiffness at the column base was induced by loosening the anchor bolts of Column 1A.
Cases 2--4 involved subsequently loosening the anchor bolts for Columns 2A, 1B, and 2B, respectively.
In Case 5, the anchor bolts of all columns were re-tightened to recreate an intact state.

\begin{table}[t]
    \centering
    \caption{Considered damage cases in the experimental validation.}\label{tab:cases}
    \fontsize{10truept}{11truept}\selectfont
    \begin{tabular*}{\textwidth}{@{\extracolsep{\fill}}rl}
        \toprule
        Damage case & Column bases with loosened anchor bolts \\
        \midrule
        0 & (intact)         \\
        1 & 1A               \\
        2 & 1A, 2A           \\
        3 & 1A, 2A, 1B       \\
        4 & 1A, 2A, 1B, 2B   \\
        5 & all re-tightened \\
        \bottomrule
    \end{tabular*}
\end{table}

\subsection{Structural Model}

In the subsequent validation, the planar frame along the A axis, referred to as Frame A, was selected for model updating (as depicted in Fig.~6(a)).
Fig.~7 showcases a structural model of this frame, incorporating two rotational springs to represent the semi-rigid column bases, with rotational stiffnesses denoted by $k_1, k_2$ (kNm/rad).
Similar to the numerical example, these stiffness values were converted into fixity factors, represented as $\gamma_1, \gamma_2$, in the Bayesian inference implementation.
The beam-column connections were treated as rigid, and the shear deformation of the elements was disregarded.
Given the interaction between the H-beams and the RC slab, the bending stiffness of the composite beam, $EI_\mathrm{cs}$, was assumed to be significantly higher than that of the H-beam, $EI_\mathrm{b}$, and expressed as $EI_\mathrm{cs} = 10^{\gamma_3} EI_\mathrm{b}$.
Hereinafter, $\gamma_3 ~ (>0)$ is termed as the \textit{beam rigidification coefficient}.
A lumped mass matrix was considered, including an additional mass for the floor slab, denoted as $m$ (kg).
Considering that the mass of the upper part of the frame was nominally 14056 kg, the nominal (target) value of $m$ was set to $m = 7000$ (kg), based on the assumption that the planar frames along the A- and B-axes (Frames A and B) equally share inertial loads.
Therefore, the uncertain structural parameters to be updated are $\bm{\theta}_\mathrm{K} = \{ \gamma_1, \gamma_2, \gamma_3 \}^\top$ and $\bm{\theta}_\mathrm{M} = \{m\}^\top$, with their nominal values presumed unknown for subsequent inference.
Additional material parameters and mechanical properties were assumed known a priori. 

\begin{figure}[!t]
    \centering
    \includegraphics[width=80truemm]{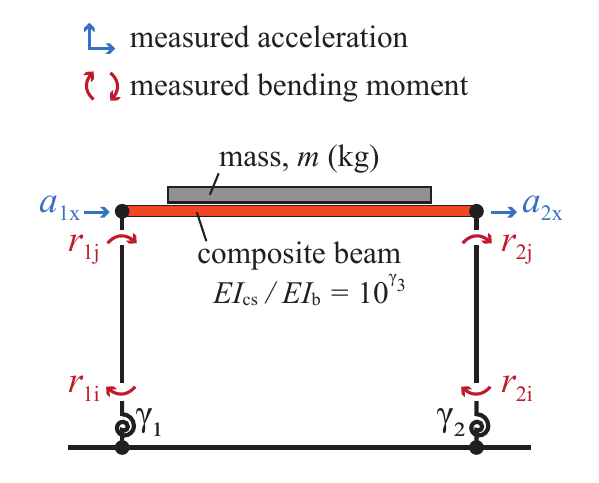}
    \caption{A planar frame assumed for the target structure in the experimental validation.}
    \label{fig:exp-model}
\end{figure}

\subsection{Modal Identification}

\begin{figure}[!t]
    \centering
    \includegraphics[width=120truemm]{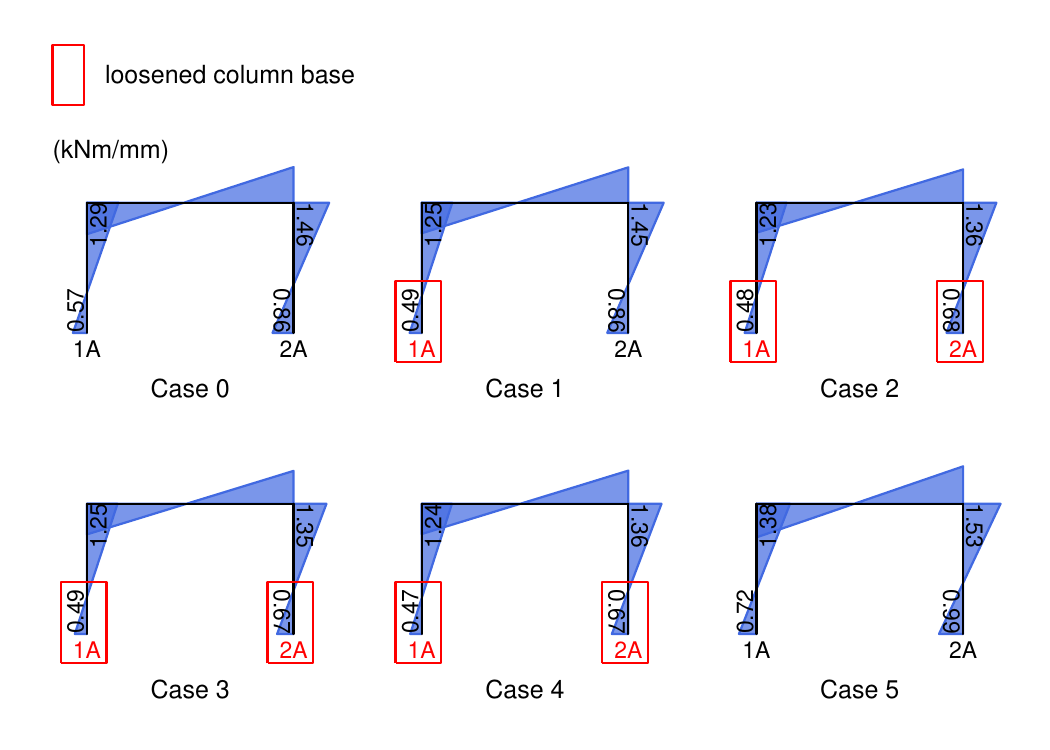}
    \caption{Identified modal bending moment distribution for all cases.}
    \label{fig:exp-mmnt}
\end{figure}

System identification applied to the strain and acceleration measurements for each damage case provided the modal properties for the $k$th mode, namely the natural angular frequency $\bar{\omega}_k$, modal acceleration (MA) $\bar{\bm{a}}_k$, and strain mode shapes (referred to as \textit{modal strain}, MS) $\bar{\bm{\varepsilon}}_k$.
For output-only system identification, stochastic realization theory, a form of the 4SID method, was employed (refer to Lemma 7.9 in \citet{katayamaT2005} or \citet{yaoyamaT2023} for an in-depth explanation). 
The critical hyperparameters---the row number of the block Hankel matrix and the system order---were set at $30$ and $6$, respectively.
Among the three successfully identified modes, i.e., EW- and NS-direction translational modes and the planar rotational mode, only the first mode (the EW-direction translational mode) was utilized in parameter estimation, focusing on the EW-direction planar frame.
In contrast to the numerical example, in the experimental validation, wireless sensor networks may induce reduced synchronization accuracy, thereby increasing the noise levels in the phase of identified complex mode shapes.
To deal with this, the absolute values of the mode shapes, with appropriate signs, were employed after verifying that the elements in the identified mode shapes exhibited nearly identical or opposite phases.
Given that the MA values in the EW direction at the roof purlin ends might not precisely correspond to those at the beam ends in Frame A owing to the in-plane rotational behavior, they were adjusted using the MA values in the NS direction.

The identified MS values at the east- and west-side edges of a cross-section, denoted by $\bar{\varepsilon}_\mathrm{E}, \bar{\varepsilon}_\mathrm{W}$, were then transformed into MBM values \citep{iyamaJ2021,iyamaJ2023,yaoyamaT2023}
\begin{linenomath*}\begin{align}
    \bar{r} &= (\bar{\varepsilon}_\mathrm{E} - \bar{\varepsilon}_\mathrm{W}) E Z / 2
\end{align}\end{linenomath*}
where Z denotes the sectional modulus. Linear interpolation using the MBM values at both the upper and lower cross-sections yielded the MBM values at both the top and base of each column.

The aforementioned procedure resulted in a dataset $\mathcal{D}_i = \{{}_i\bar{\omega}_1, {}_i\bar{\bm{d}}_1, {}_i\bar{\bm{r}}_1\}$ for Case $i = 0,...,5$, where ${}_i\bar{\bm{d}}_1 \in \mathbb{R}^2$ consists of the lateral MD values at both ends of the beam and ${}_i\bar{\bm{r}}_1 \in \mathbb{R}^4$ contains the MBM values at the top and bottom of both two columns, as depicted in Fig.~7.
Both the MD and MBM vectors were normalized such that the MD vector satisfies $\|{}_i\bar{\bm{d}}_1\|_2 = 1$, preserving the relationship between ${}_i\bar{\bm{d}}_1$ and ${}_i\bar{\bm{r}}_1$.
The MBM values for Cases 0 to 5 are presented in Fig.~8.
The MBM value at the base of Column 1A significantly decreases from Case 0 to 1 and increases from Case 4 to 5, corresponding to the loosening and re-tightening of the anchor bolts of Column 1A.
The same trend is observed for the base of Column 2A.
These results imply that the proposed stress-resultant-based approach is sufficiently sensitive to localized structural damages.

\subsection{Bayesian Inference}

\begin{figure}[!t]
    \centering
    \includegraphics[width=120truemm]{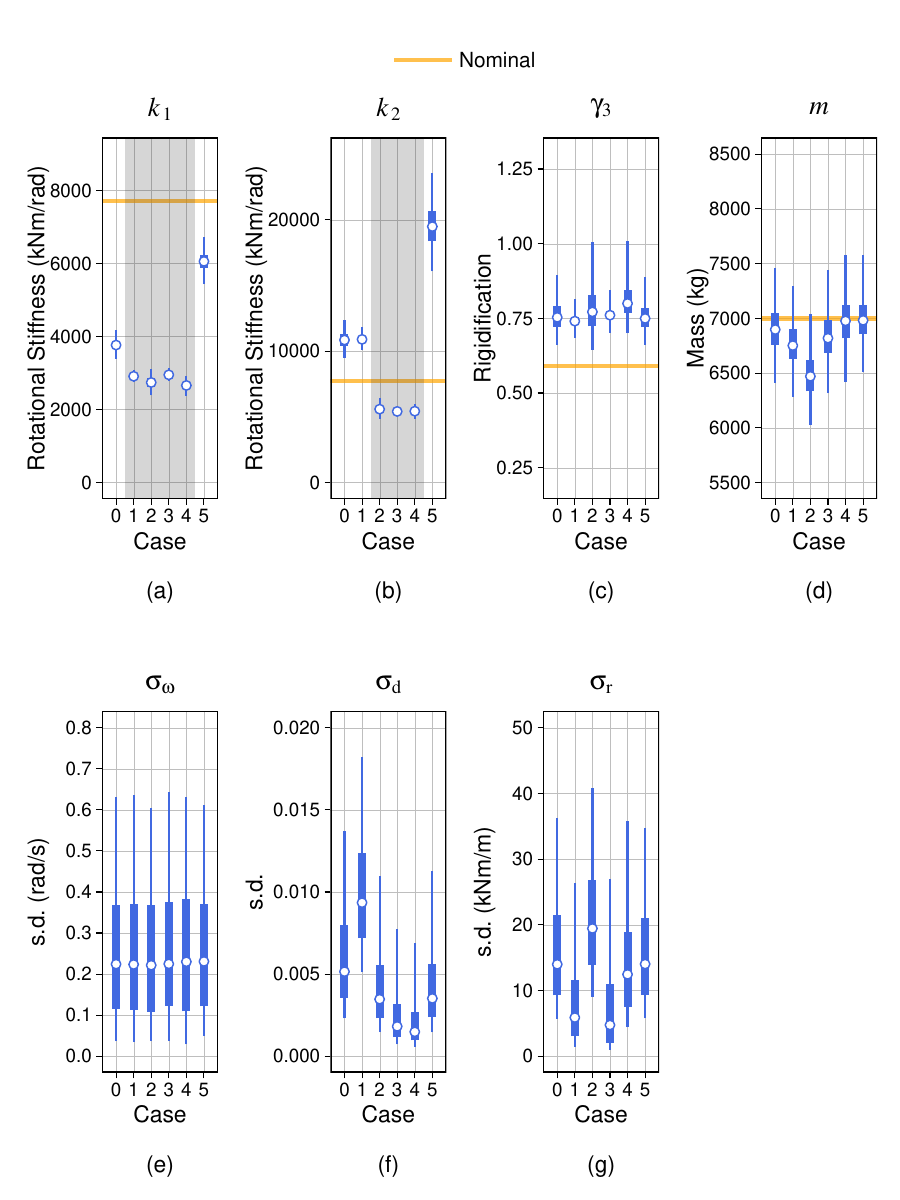}
    \caption{Distributions of the posterior samples for all cases: (a)--(b) stiffness parameters; (c) beam rigidification coefficient; (d) mass parameter; (e)--(g) statistical parameters for the natural frequency, modal displacement, and modal bending moment.}
    \label{fig:exp-posterior}
\end{figure}

\begin{figure}[!t]
    \centering
    \includegraphics[width=\textwidth]{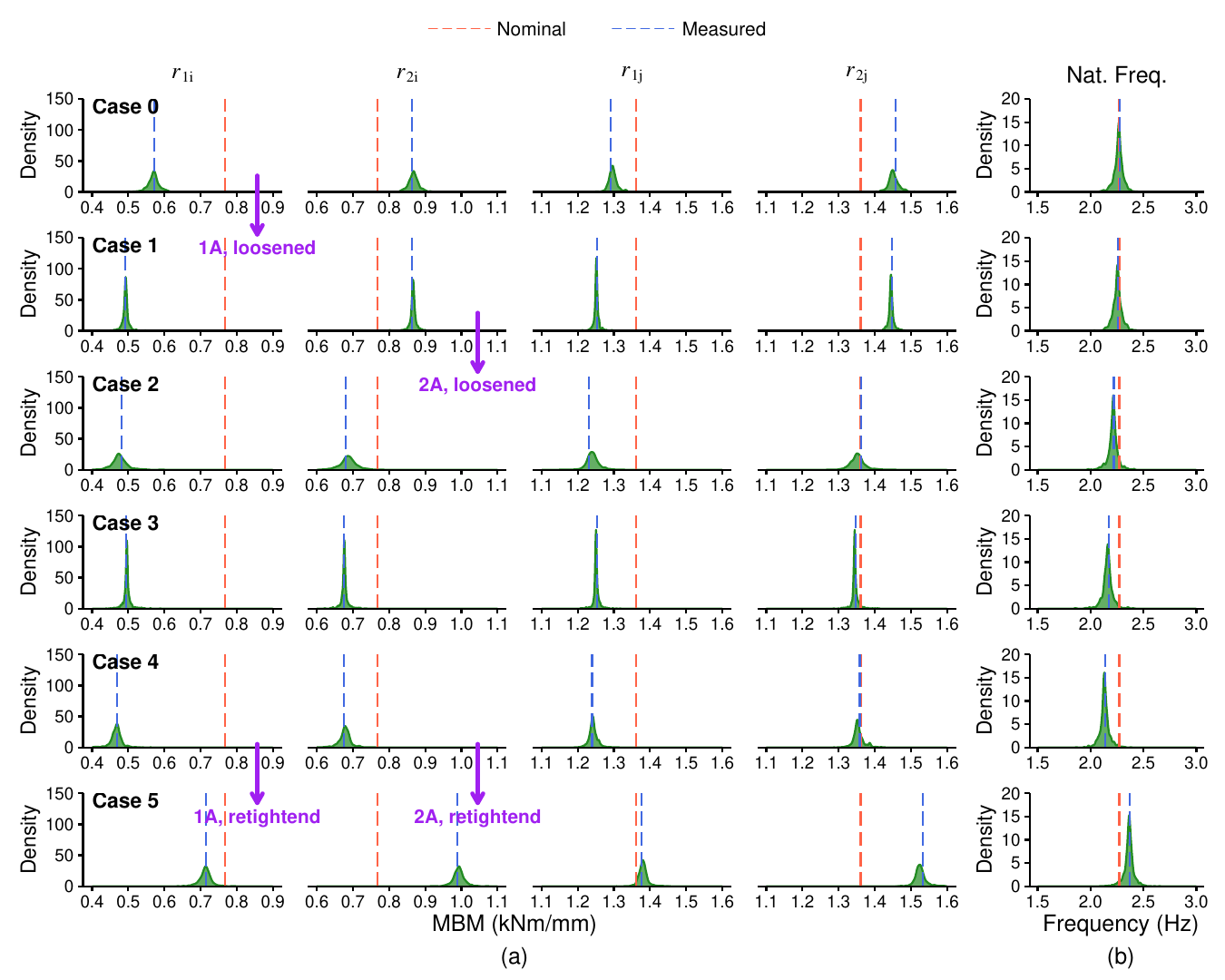}
    \caption{Predictive PDFs in comparison with the nominal and measured values for Case 0--5: (a) modal bending moment at the bottom and top of Columns 1A and 2A ($r_\mathrm{1i},r_\mathrm{1j},r_\mathrm{2i},r_\mathrm{2j}$); (b) natural frequency.}
    \label{fig:exp-prediction}
\end{figure}

Similar to the previous numerical example, the prior PDFs for the fixity factors $\gamma_1,\gamma_2$ were defined as noninformative uniform priors, $\mathcal{U}(0, 1)$.
The prior PDFs for the beam rigidification coefficient $\gamma_3$ and mass $m$ (kg) were also set as uniform distributions over a sufficiently wide range, $\gamma_3 \sim \mathcal{U}(0, 3)$ and $m \sim \mathcal{U}(0, 10^5)$, to mitigate potential biases.
The statistical parameters $\sigma_\mathrm{d}$, $\sigma_\mathrm{r}$, and $\sigma_\omega$ adopt the weakly informative priors in Eq.~(\ref{eq:weakinfo-d})--(\ref{eq:weakinfo-w}), where $s_\mathrm{d} = 0.01$, $s_\mathrm{r} = 20$ (kNm/m), and $s_\omega = 0.1 \pi$ (rad/s).
The posterior PDFs were estimated using the NUTS algorithm and Stan implementations, maintaining the same number of chains, sample size, and burn-in period as in the previous numerical example.
The convergence of all parameters was confirmed with the metric $\widehat{R} < 1.1$.

Fig.~9 displays the posterior sample distributions for all parameters, where the fixity factors $\gamma_1, \gamma_2$ are converted into rotational stiffness values $k_1, k_2$ (kNm/rad).
The orange horizontal lines in Fig.~9(a)--(c) indicate the nominal values based on AIJ recommendations \citep{aij2021,aij2010}: 7722 (kNm/rad) for $k_1, k_2$ and 0.59 ($EI_\mathrm{cs}/EI_\mathrm{b} = 3.89$) for $\gamma_3$.
In Fig.~9(d), the orange line represents the nominal (target) mass, $m = 7000$ (kg).
The median of $k_1$ significantly decreases from Case 0 to 1, remains relatively stable between Cases 1 and 4, and increases considerably from Case 4 to 5.
Notably, even the 5th percentile in Case 1 is lower than the 95th percentile in Case 0, and the 95th percentile in Case 5 is substantially higher than the 5th percentile in Case 4, whereas the 90\% credible intervals in Cases 1--4 overlap.
These observations accurately reflect the simulated damage and recovery scenarios: loosening of the anchor bolts of column 1A between Cases 0 and 1, and re-tightening the anchor bolts of all columns between Cases 4 and 5.
A similar pattern is noted for $k_2$.
Despite the medians of $k_1$ and $k_2$ being in the same order of magnitude as the nominal values, there is a noticeable divergence from these nominal values and between each other across all cases.
This observation underscores the importance of model updating to accurately represent column-base-specific structural performance, even in the absence of structural damage.
The median value of the rigidification factor $\gamma_3$ remains consistently around 0.75 (i.e., $EI_\mathrm{cs}/EI_\mathrm{b} \simeq 5.6$) across all cases, unaffected by changes in the rotational stiffness of the column bases.
The median of $m$ stays close to the nominal (target) value.
The lower values in Cases 1--2 might be attributed to the slightly reduced participation of Frame A in bearing inertial force compared to Frame B, owing to the decreased stiffness at the column bases.

Figure~10 presents the predictive PDFs of (a) the MBM values at the bottom and top of the two columns ($r_\mathrm{1i}, ..., r_\mathrm{2j}$ in Fig.~7) and (b) the natural frequency.
These densities were derived using kernel density estimation (KDE) applied to the MBM value samples simulated for the posterior samples of structural parameters.
The nominal and measured values are represented by blue and red vertical dashed lines, respectively.
Despite certain discrepancies between nominal and measured values in all cases, the peaks in the predictive PDFs closely correspond with the measured values, signifying enhanced response predictions. 

In Cases 0--2, the simulated damage to the column bases in Frame A led to reduced MBM values at these column bases; however, their impact on the natural frequency was relatively minor due to the preserved stiffness of Frame B.
This resulted in a reduction in mass, indicating a lesser inertial force borne by Frame A, as depicted in Fig.~9(d).
Conversely, from Case 2 to 4, the simulated damages in Frame B minimally affected the MBM values in Frame A and only led to a slight decrease in the natural frequency, reflecting the ``recovery'' of the inertial force borne by Frame A.
These findings demonstrate that the stress-resultant-based approach effectively discerns elemental (local) stiffness from the global stiffness of the entire structural system.
Consequently, the proposed method enhances pre-event response predictions and post-event damage localization.

\section{Conclusions}

This study introduces a stress-resultant-based methodology for Bayesian model updating of frame structures, independent of mass assumptions, to improve various levels of SHM, including damage detection, localization, quantification, and response prediction.
The Bayesian approach integrates strain and acceleration measurements, evaluating two types of likelihood functions: one based on modal analysis and the other on static analysis.
The static-analysis-based likelihood leverages the relationship between external deformation, represented by modal displacement, and internal force, indicated by modal stress resultant, thereby eliminating the need for prior assumptions regarding external forces and mass parameters.
We presented a numerical example of a two-story planar frame, featuring rotational springs at the beam ends and column bases, with unknown mass parameters and spring stiffnesses.
Despite observing minor residuals potentially due to biases in system identification, the approach effectively estimated both mass and stiffness parameters with reduced and quantified uncertainties.
Furthermore, it reliably predicted peak values of acceleration and bending moment responses to ``future'' ground motion.
These outcomes remained consistent across varying levels of observation noise and different realizations of noise and ground motion, underscoring the robustness of the proposed method.
Experimental validation involved a full-scale, one-story, moment-resisting steel frame, with localized structural damage at the column bases simulated by loosening the anchor bolts.
The approach appropriately estimated the mass of the frame with reasonable uncertainties and quantitatively assessed the stiffness decrease (and subsequent increase) of the column bases, demonstrating a high sensitivity to localized structural damage.
These findings signify that the proposed stress-resultant-based Bayesian approach facilitates the simultaneous estimation of mass and stiffness parameters in a probabilistic manner, thereby enhancing both post-event damage assessment and pre-event response prediction in existing civil structures.

Future research directions may include:
(1) adopting Bayesian methods of system identification for addressing the bias introduced when estimating modal stress resultant;
(2) additional experimental validation focusing on structural response to ground motion;
\textcolor{black}{
(3) a more thorough investigation into the applicability of the proposed approach, such as examining the effects of different sensor configurations on model updating performance; and
(4) examining the efficiency of (and potential improvements to) the proposed approach for high-dimensional model updating problems.
}

\section{Data Availability Statement}

Some or all data, models, or code that support the findings of this study are available from the corresponding author upon reasonable request.

\section{Acknowledgments}

The dynamic loading test in this study was part of a research conducted in collaboration with Prof. Satoshi Yamada and Prof. Tsuyoshi Seike at the University of Tokyo and was supported by JSPS KAKENHI (Grant-in-Aid for Scientific Research) (B) Grant Number JP20H02293.

\appendix

\section{Reduction and recovery of stiffness matrix}

Referring to \citet{guyanRJ1965}, the reduction (and recovery) of stiffness matrix $\mathbf{K}$ is formulated as follows.
The nodal displacement vector $\bm{d}$ is rearranged and decomposed into two groups: $\bm{d} = \{\bm{d}_\mathrm{m}^\top, \bm{d}_\mathrm{s}^\top \}^\top$, where $\bm{d}_\mathrm{m} \in \mathbb{R}^{N_\mathrm{m}}$ denotes responses at ``master'' DOFs, and $\bm{d}_\mathrm{s} \in \mathbb{R}^{N_\mathrm{s}}$ denotes responses at ``slave'' DOFs for which external forces are negligible, i.e., almost no masses are assigned.
Accordingly, the global stiffness matrix $\mathbf{K}$ is rearranged and decomposed as follows.
\begin{linenomath*}\begin{align}
    \begin{bmatrix}
        \mathbf{K}_\mathrm{mm} & \mathbf{K}_\mathrm{sm} \\
        \mathbf{K}_\mathrm{ms} & \mathbf{K}_\mathrm{ss}
    \end{bmatrix}
    \begin{Bmatrix}
        \bm{d}_\mathrm{m} \\
        \bm{d}_\mathrm{s}
    \end{Bmatrix}
    =
    \begin{Bmatrix}
        \bm{f}_\mathrm{m} \\
        \bm{f}_\mathrm{s}
    \end{Bmatrix}
\end{align}\end{linenomath*}
where $\bm{f}_\mathrm{m}$ and $\bm{f}_\mathrm{s}$ denote external forces applied at the master and slave DOFs.
Assuming $\bm{f}_\mathrm{s} = \mathbf{0}$, the following reduced form is obtained:
\begin{linenomath*}\begin{align}\label{eq:reduced}
    \bm{f}_\mathrm{m} = \left( \mathbf{K}_\mathrm{mm} - \mathbf{K}_\mathrm{sm} \mathbf{K}_\mathrm{ss}^{-1} \mathbf{K}_\mathrm{ms} \right) \bm{d}_\mathrm{m}
\end{align}\end{linenomath*}
The recovery of a complete vector for all DOFs can be expressed as
\begin{linenomath*}\begin{align}
    \bm{d} = \mathbf{\Gamma} \bm{d}_\mathrm{m}
\end{align}\end{linenomath*}
where
\begin{linenomath*}\begin{align}
    \mathbf{\Gamma} =
    \begin{bmatrix}
        \mathbf{I}_{N_\mathrm{m}} \\
        - \mathbf{K}_\mathrm{ss}^{-1} \mathbf{K}_\mathrm{ms}
    \end{bmatrix}
\end{align}\end{linenomath*}
where $\mathbf{I}_{N_\mathrm{m}} \in \mathbb{R}^{N_\mathrm{m}\times N_\mathrm{m}}$ denotes the identity matrix.
In this study, all master nodal DOFs are assumed to be measured.
Thus, $N_\mathrm{m} = N_\mathrm{obs}$.
The reduced form in Eq.~(\ref{eq:reduced}) is exploited for modal analysis, and $\mathbf{\Gamma}$ defined above serves as $\mathbf{\Gamma}(\bm{\theta}_\mathrm{K})$ in Eq.~(\ref{eq:static_anal}).

\bibliographystyle{elsarticle-harv} 
\bibliography{2308_asce}

\end{document}